\documentclass[a4paper]{VisionStyle}
\usepackage{epsfig}

\begin{document}

\title{Chandra Deep X-ray Observation on the Galactic Plane}

\author{K.\ Ebisawa\inst{1,2,3},   A. Bamba\inst{4}, H.\  Kaneda\inst{5},  Y.\ Maeda\inst{5},  
A.\ Paizis\inst{1}, G.\ Sato\inst{5} and  S.\ Yamauchi\inst{6}} 
\institute{
  INTEGRAL Science Data Center, Chemin d'Ecogia 16, Versoix, Switzerland
\and 
  code 662, NASA/GSFC, Greenbelt, MD 20771, USA
\and
  Universities Space Research Association
\and 
  Kyoto University, Sakyo-ku, Kyoto, 606-8502, Japan
\and
  Institute of Space and Astronautical Science, Yoshinodai, 
  Sagamihara, Kanagawa 229-8510, Japan
\and
  Iwate University, Iwate, 020-8550, Japan
}

\maketitle 

\begin{abstract}

Using the Chandra ACIS-I instruments, 
we have carried out the deepest X-ray observation on a typical Galactic plane 
region at  $l \approx 28.^\circ5$, where no discrete X-ray sources have been
known previously.  We have detected, as well as strong diffuse emission, 
275 new point X-ray sources (4 $\sigma$ confidence) 
within two partially overlapping fields ($\sim $250 arcmin$^2$ in total)
down to $\sim 3 \times 10^{-15} $ 
erg s$^{-1}$ cm$^{-2}$
(2 -- 10 keV) 
or   $\sim 7 \times 10^{-16} $ erg s$^{-1}$ cm$^{-2}$ (0.5 -- 2 keV).
We have studied spectral distribution of these point sources, and 
found that very soft sources  detected only below
$\sim 3 $ keV are more numerous than hard sources detected only above $\sim$ 3 keV.
Only small number of sources are detected both in the soft and hard bands.
Surface density of the hard sources is almost consistent with that at high Galactic regions,
thus most of the hard sources are considered to be Active Galactic Nuclei seen through
the milky way.  On the other hand, some of the bright hard X-ray sources 
which show extremely flat spectra and iron line or edge features are considered
to be  Galactic, presumably quiescent dwarf novae.
The soft sources show thermal spectra and small interstellar hydrogen column densities,
and some of them exhibit X-ray flares.  Therefore, most of the 
soft sources are probably X-ray active nearby late type stars.

\end{abstract}

\section{Introduction}
  
The Galactic plane has been known to be  a strong hard X-ray (2 -- 10keV) emitter for nearly
20 years  (e.g., Worrall et al.\ 1982; Warwick et al.\ 1985; Koyama et al.\ 1986).
The emission forms a narrow continuous ridge, thus it is often called
the Galactic Ridge X-ray Emission (GRXE).
GRXE exhibits emission lines from highly 
ionized heavy elements  such as Si, S and Fe,  which suggests that GRXE is originated from
thin hot plasmas with a temperature of several keV (Koyama et al.\ 1986).
However, whether  GRXE is composed of
numerous point sources or  truly diffuse emission  has been an unsolved problem,
mostly because previous instruments have not had good enough spatial resolution 
in hard X-ray band ($>$ 2 keV).
The Chandra X-ray mirror has a superior angular resolution ($\sim 0.5''$),
which allows one to distinguish numerous dim point sources and truly diffuse X-ray emission.
Therefore, we planned  Chandra observation on a typical Galactic plane region to resolve origin
of the hard X-ray emission from the Galactic plane.

\begin{figure*}[ht]
  \begin{center}
    \epsfig{file=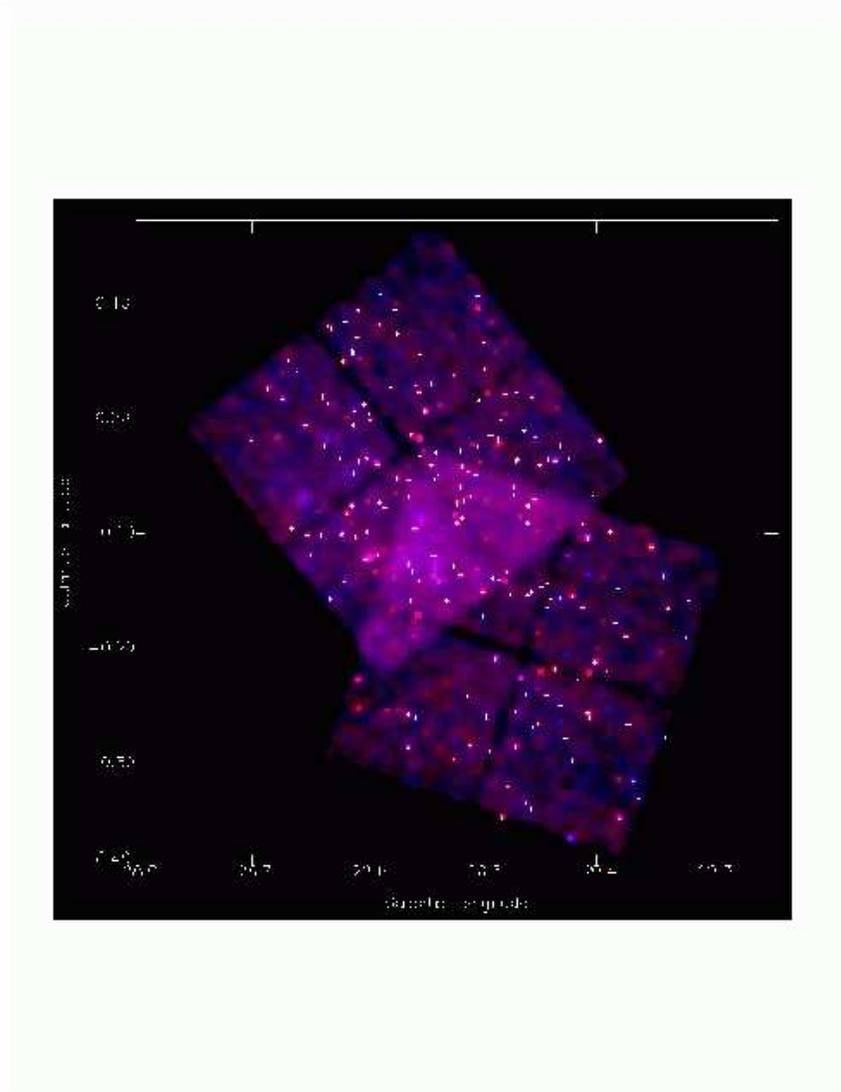, width=14cm}
  \end{center}
\caption{Superposed image of the  two Chandra observations (exposure is not
corrected). South is AO1 
and North is the AO2 fields, each for 100 ksec exposure. Hard X-rays in 3 -- 8 keV
are expressed in blue, and soft X-rays in 0.5 -- 3 keV are shown in red.  The 275 detected point
sources are shown in crosses.}  
\label{fauthor-E1_fig:fig1}
\end{figure*}

\begin{figure*}[ht]
  \begin{center}
    \epsfig{file=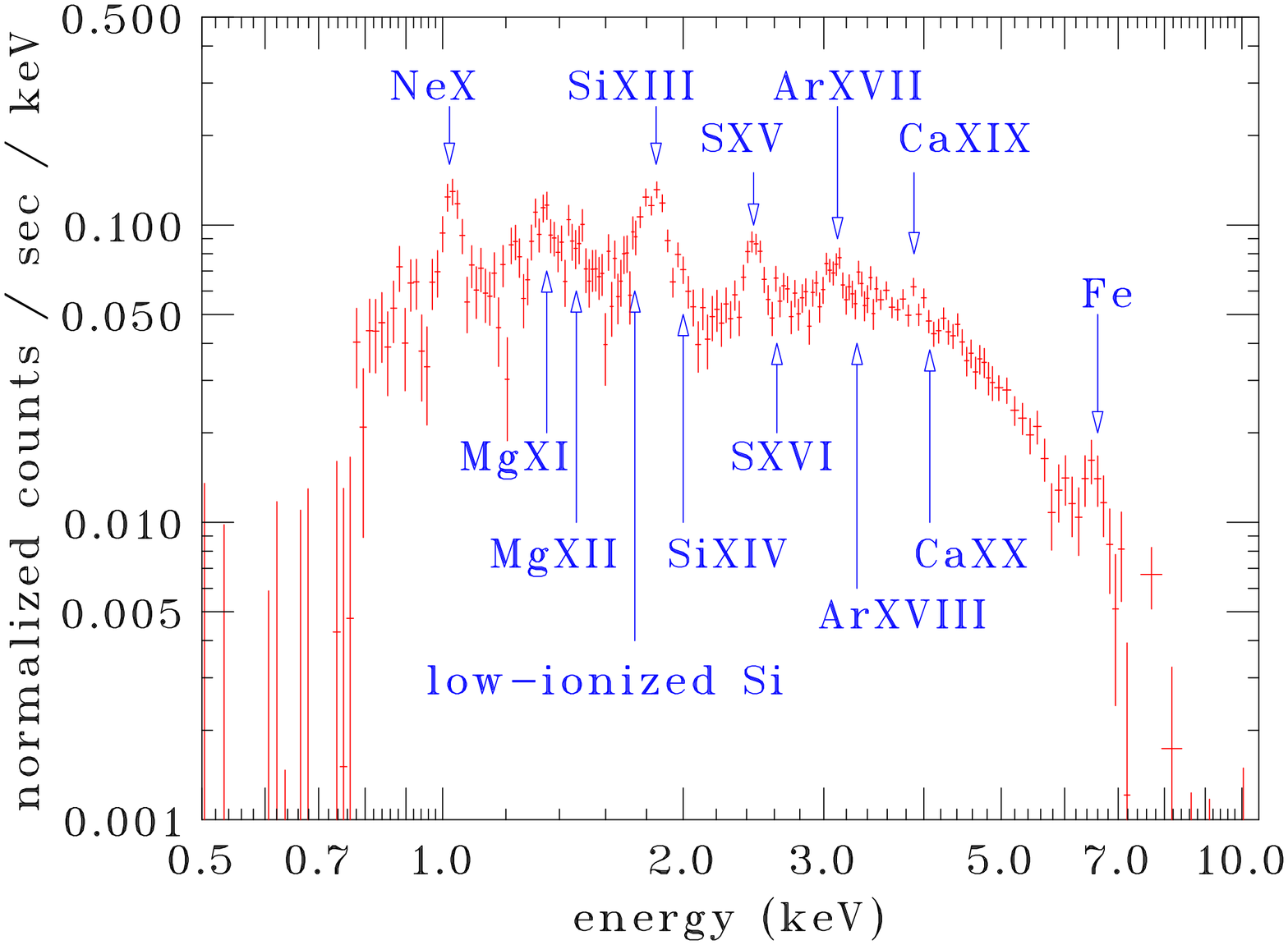, angle=90,width=14cm}
  \end{center}
\caption{Energy spectra of the total X-rays in our fields.}  
\label{fauthor-E1_fig:fig1}
\end{figure*}

\section{Observation}
Using Chandra ACIS-I, we observed a  region around
$(l,b) \approx (+28^\circ.45,-0^\circ.2)$,
where the ASCA satellite  could not  find any point sources 
brighter than $\sim 2 \times 10^{-13}$ erg s$^{-1}$ cm$^{-2}$ (2 -- 10 keV)
(Yamauchi et al.\ 1996; Kaneda et al.\ 1997).  We carried out two pointings,
each for 100 ksec in AO1 and AO2, with slightly overlapping 
the fields (Figure 1).  Total area of the observed field is $\sim 250 $ arcmin$^2$.
The first result from the AO1 result was published in Ebisawa et al.\ (2001).

\section{Data Analysis and Results}

\subsection{Point Source Search}
In our fields, we see many point sources as well as strong diffuse emission.
We have carried out point source search using the ``wavdetect'' program
in the CIAO data analysis package.  We have searched for sources in the
0.5 -- 3 keV, 3 -- 8 keV and 0.5 -- 8 keV independently.  
The sources which exceed 4 $\sigma$ significance  either in the
three energy bands are considered as the true detection.  
On the AO1 and AO2 overlapping field, we searched for
sources for AO1 and AO2 separately, and added the two significances 
quadratically.

Thus, we have detected  275 new  X-ray point sources  within the AO1 and AO2  field of view.
In the soft band, 183 sources have been detected, while in the hard band we have detected
79 sources.  Note that only 26 sources have been detected both in the soft and
hard bands.  Our sensitivity is $ \sim 3 \times 10^{-15}$  erg s$^{-1}$ cm$^{-2}$ (2 -- 10 keV)
and $ \sim 7 \times 10^{-16}$  erg s$^{-1}$ cm$^{-2}$ (0.5 -- 2 keV).

Four soft sources at $(\alpha,\delta)=(280.^\circ99, -3.^\circ91)$ actually
seem to compose an extended feature, that is probably a blob in the 
super nova remnant AX J 1843.8--0352/G28.6--0.1 (Koyama et al. 2002).
We tried to identify other  new X-ray point sources with those in published catalogs.
About a dozen of the soft sources have been identified in the United States
Naval Observatory A2.0 catalog with the R magnitude brighter than $\sim$ 18.
On the other hand, no hard sources have been optically identified.
Deep optical or near infrared follow-up observations are anticipated
to identify some of these  hard X-ray sources\footnote{We have been
granted two nights to observe this field in the near-infrared band 
with the ESO/NTT SOFI.}.

\subsection{Origin of the Galactic Hard X-ray Emission}

After subtracting the expected non X-ray background, we calculated the total hard X-ray energy
flux from our field of view, that was $\sim 1.1 \times 10^{-10} $erg s$^{-1}$  cm$^{-2}$ (2 -- 10 keV).
On the other hand, total hard X-ray flux calculated by combining all the point sources was 
$\sim 10^{-11} $erg s$^{-1}$  cm$^2$ (2 -- 10 keV); namely point sources
account for  only $\sim 10 \% $ of the total hard X-ray flux in the field of view.
We compared  surface density of the hard point sources in 
our field view with those at high Galactic blank fields (e.g. Giacconi 2001).
After the Galactic absorption ($\sim 6 \times 10^{21}$ cm$^{-2}$) is taken into account,
we found the number density of the hard X-ray sources 
in our Galactic plane field is comparable to those in high Galactic bland fields
(Ebisawa et al.\ 2001). 
Therefore, we concluded that most of  the point sources are   extragalactic,
presumably   active galaxies  seen through the Galactic disk.
Present result clearly indicates that GRXE is diffuse origin,
which indicates omnipresence of the energetic plasma along the Galactic plane.

\subsection{Diffuse Emission Energy Spectra}
We have made energy spectra of the Galactic X-rays (diffuse emission
and Galactic point sources) by subtracting the non X-ray background
and contribution of the extragalactic component.  Before subtracting
the background, we performed the CTI correction on our event lists 
in order  to
improve the energy resolution.
We took the extragalactic observations on HDF-N (sequence \# 90030 and 900061), whose
exposure time is 337 ksec in total.  We subtracted the energy spectrum of the HDF-N from
that of our fields, by adjusting the normalization of the former so that 
 the instrumental Ni K$_\alpha$ lines in the two energy spectra cancel each other.  The resultant
energy spectrum is shown in Figure 2.  Numerous emission lines expected from
highly ionized plasma are clearly seen.

\subsection{Spectral Distribution of Point Sources}

In order to study spectral distribution of the point sources,
we calculated the spectral hardness ratio (HR) for individual sources.
We corrected for positional difference of the instrumental response and difference
of the exposure time, such that the corrected count rates are those
expected when source are located at the ACIS-I aim point and observed 
for 100 ksec exposure.
We defined the hardness ratio as $ HR \equiv (H-S)/(H+S)$, where $H$ is the corrected
count rate in the hard energy band (3 -- 8 keV), and $S$ is that in the soft energy band
(0.5 -- 2 keV).  In figure 3, we show histogram of the spectral distribution of the point sources
(left),
and the  counting rate vs.\  HR ratio diagram (right).

From figure 3 (left), we can see that the softest sources with $-1 \le HR \le -0.8$ are most numerous.
The number of sources decreases as HR increases, but it increases again 
as HR exceeds  0.6.  If we see the counting rate vs.\ HR diagram (figure 3, right),
relatively dim sources with the corrected count rates $\le$ 100 cts/100 ksec 
seem to be clustered into two distinct classes, very soft sources with $HR \le -0.8$
and very hard sources with $HR \ge 0.6$.  Brighter sources with the
 corrected count rates $\ge$ 100 cts/100 ksec do not seem to have such a tendency.

\begin{figure*}[ht]
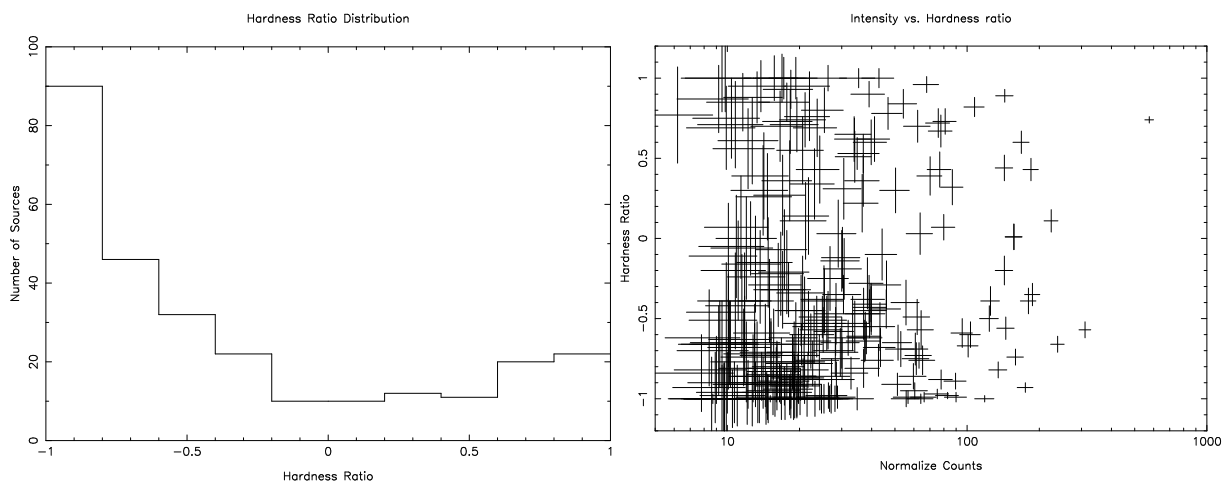

  \begin{center}
\epsfig{file=kebisawa_E1_fig3_left.eps,angle=270,width=8cm}
\epsfig{file=kebisawa_E1_fig3_right.eps,angle=270,width=8cm}
  \end{center}
\caption{Histogram of the hardness ratio distribution of the sources
(left) and the count rate vs.\  hardness ratio diagram (right).}  
\end{figure*}

\subsection{Point Source Energy Spectra}

Individual sources are too dim to make energy spectra, so we combined
sources with similar $HR$ to make a single energy spectrum to study
characteristics of the point sources as a spectral class.
We combined 91 soft sources with $HR \le -0.8$ and 44 hard sources with
$HR \ge 0.6$, respectively.  In Figure 4, we show these energy spectra.

The combined soft spectrum may be roughly fitted with a thin thermal plasma
emission model with $kT  = 0.76\pm0.04$  keV with the interstellar
hydrogen column density $N_H =
(1.0\pm 0.1) \times  10^{22}$  cm$^{-2}$.  The small column density
relative to the Galactic value ($\sim 6 \times  10^{22}$  cm$^{-2}$)  suggests these are nearby sources,
and the low temperature thermal spectra suggests they are X-ray active stars.
On the other hand, the hard spectrum may be fitted with a power-law 
with the index $1.2 \pm 0.4$, and $N_H = (8.0\pm 2.4) \times  10^{22}$  cm$^{-2}$.
The large column density and the flat spectrum are consistent with the idea
that most of the hard sources are extragalactic AGN.  However, the 
column density is slightly higher than the Galactic value, and the spectral
slope is also slightly flatter than that expected from the 
composition of dim AGN composing cosmic X-ray background (1.4).
These facts, as well as the hint of iron K-line emission or edge
feature in the composite energy spectrum (figure 4, right), suggests
there are several Galactic hard X-ray sources which have flat spectra
and iron emission or edge feature. 
Quiescent dwarf novae are likely candidates for the Galactic hard X-ray sources
with such spectral characteristics (Mukai and Shiokawa 1993; Watson 1999).

\subsection{Point Source Time Variation}

We have studied time variation of the point sources.
For each source, we made light curves
with 3000 sec and 10000 sec bins.  For each light curve,
we performed the Kolmogorov-Smirnov test, and the source
is considered to be variable if both light curves
show the variation above the  99.9 \% significance.
Thereby, 17 sources  are found to be significantly variable.
Among them, 13 source have the $HR <0$ and 4 sources 
have $HR >0$.  Soft sources with $HR <0$  tend to show
flare-like variations.  Typical flares from some of the soft sources
are shown in figure 5.

\begin{figure*}[ht]
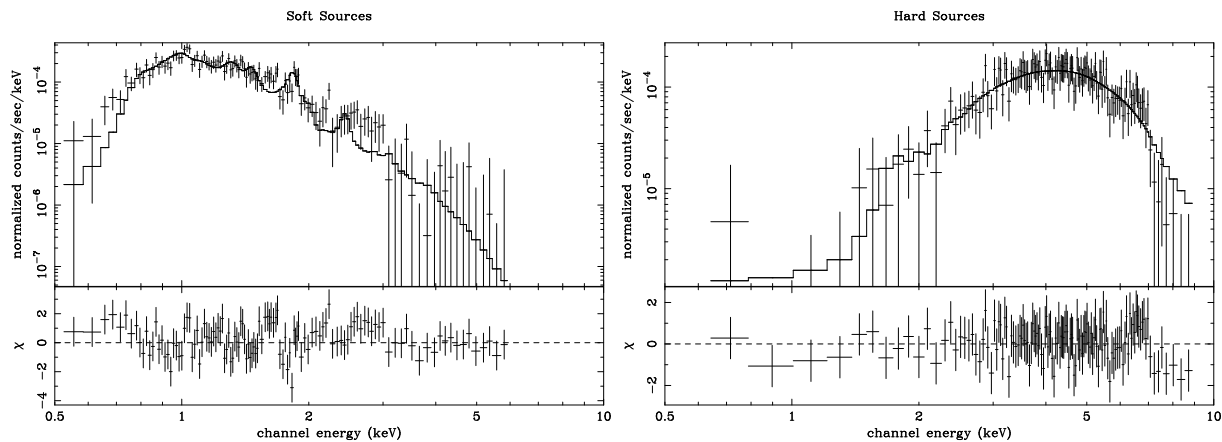

  \begin{center}
    \epsfig{file=kebisawa_E1_fig4_left.eps,angle=270,width=8cm}
    \epsfig{file=kebisawa_E1_fig4_right.eps,angle=270,width=8cm}
  \end{center}
\caption{Energy spectra of the combined soft sources (left) and hard sources (right).}  
\end{figure*}

\section{Discussion}
\subsection{Origin of the Diffuse hard X-ray emission}

We found that   GRXE has a truly diffuse origin,  then the question
is how to produce and maintain such high energetic plasma.
However, there are problems in interpreting GRXE in terms of simple equilibrium
thermal plasma.
The plasma temperature needed to explain the observed spectra,  5 -- 10 keV, 
is much higher than that can be bound by Galactic gravity (Warwick et al.\ 1985).
Also,  energy density of GRXE, $\sim$ 10 eV/cm$^3$, 
is one or two orders of magnitude 
higher than those of other constituents in the interstellar space, 
such as cosmic rays, Galactic magnetic fields, or ordinary
interstellar medium  (Koyama et al.\ 1986; Kaneda et al.\ 1997).
We do not know how to heat the plasma gas up to such a high temperature, and 
hold the hot gas within the Galactic plane.

On the other hand, strong diffuse  gamma-ray ($\sim$ 100 keV -- 1 MeV) emission
is observed from the Galactic plane (e.g., Gehrels and Tueller 1993;
Skibo et al.\ 1997), which is suggested to have a non-thermal origin,  as the energy spectrum
is represented with a power-law without a thermal cut-off.  
GRXE, besides the thermal component, also 
has a power-law hard tail component which extends above $\sim$ 10 keV
(Yamasaki et al.\ 1997; Valinia and Marshall 1998).  This hard-tail component
seems to be  smoothly connected to the gamma-ray component (Valinia et al.\ 2000a),
although their physical relationship has not been fully understood.

Currently there are no accepted theoretical models which can explain the origin
of both GRXE and gamma-ray emission.   Some argue
that interstellar magnetic field is playing a significant role to heat and
confine the hot plasma (Tanuma et al.\ 1999).
Others argue that interaction of low energy cosmic-ray electrons 
(Valinia et al.\ 2000b) or heavy  ions (Tanaka et al.\ 2000; Tanaka 2002) with 
interstellar medium is mainly responsible for GRXE and gamma-ray emission.
Different heating or acceleration mechanism of the plasma will result in different
plasma conditions, which are reflected in the emission lines.
From precise measurements of the iron emission lines, presumably performed with XMM, 
we may diagnose the plasma and will strongly constrain the origin of GRXE.

\subsection{Origin of the Point Sources}
We concluded that most of the hard X-ray point sources discovered with Chandra are extragalactic, since 
the number of hard X-ray point sources  does not
significantly exceed that expected for extragalactic sources.
From statistical arguments, we concluded that number of 
Galactic hard X-ray  point sources between the 
flux ranges $ 3 \times 10^{-15}$ and $ 2 \times 10^{-13}$ 
erg s$^{-1}$ cm$^{-2}$ (2--10 keV) does not exceed $\sim 260 $ sources/degree$^2$ (90 \% upper-limit;
Ebisawa et al.\ 2001),
while the number of extragalactic source is $\sim 560$  sources/degree$^2$ (e.g., Giacconi 
et al.\ 2001).
In other words, 
among the $\sim$ 79 point sources (4 $\sigma$ confidence) we detected
in our AO1 and AO2 Chandra fields (250 arcmin$^2$; Figure 1), 
there may be up to $\sim$ 18 Galactic sources.  
From precise X-ray spectral observations, we may tell which sources are 
Galactic ones such as quiescent dwarf novae  (Mukai and Shiokawa 1993; Watson 1999).
Also these Galactic sources are likely to be identified by near-infrared observations,
while extra galactic sources will be completely obscured  and may not 
be visible in infrared.

Soft point sources are most likely nearby X-ray active stars, as they have small
interstellar hydrogen column densities and low plasma temperatures.
They are dim in optical lights (only small fraction was identified), 
and several sources show flare like time variations; these facts
suggests the soft sources are mostly late type stars whose X-ray emission
is due to their magnetic activities.  Nature of these soft
sources will be also clarified by more precise X-ray
spectral observations and the follow-up near-infrared observations.

\begin{figure*}[ht]
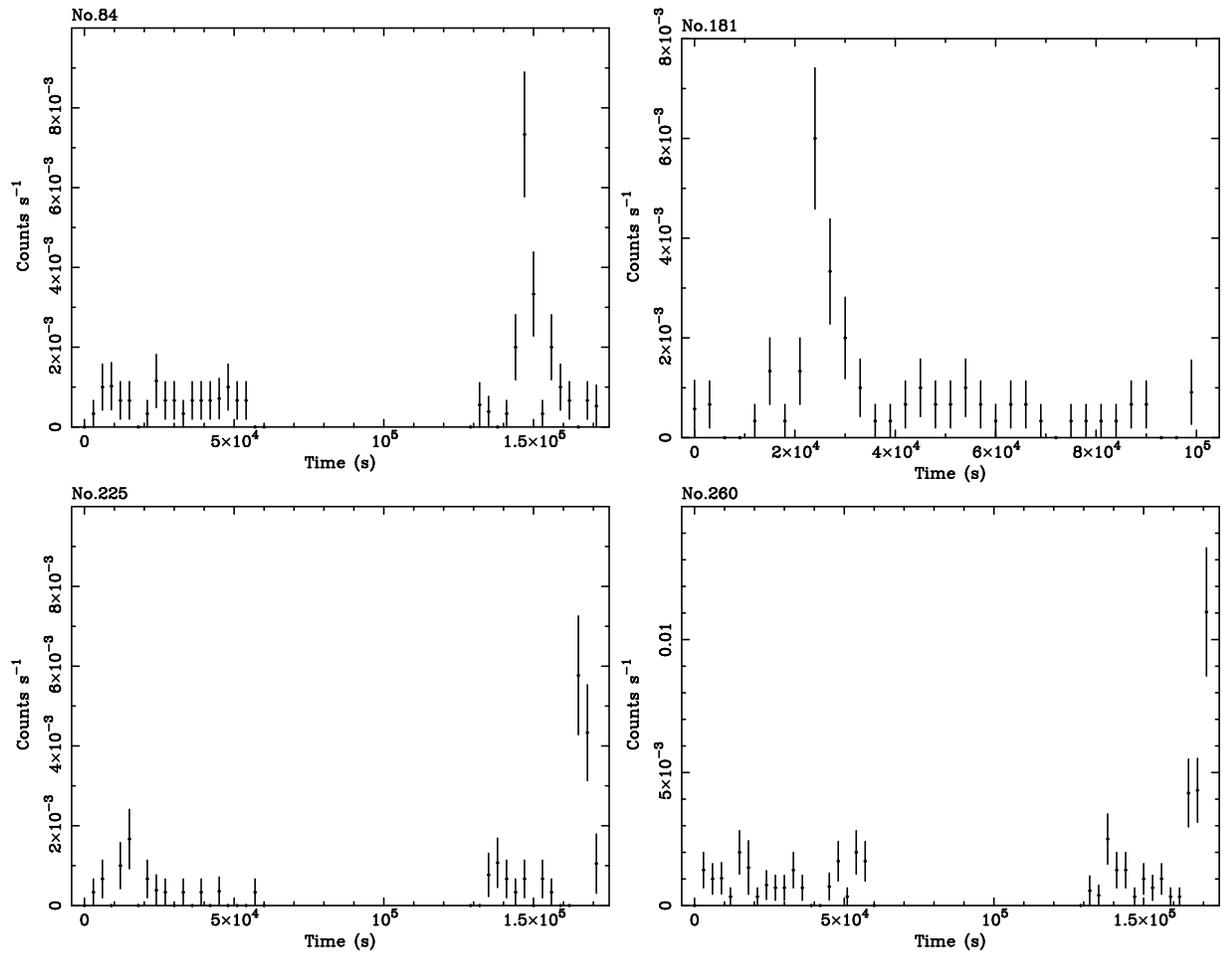

  \begin{center}
    \epsfig{file=kebisawa_E1_fig5_1.eps,angle=270,width=8cm}
    \epsfig{file=kebisawa_E1_fig5_2.eps,angle=270,width=8cm}
    \epsfig{file=kebisawa_E1_fig5_3.eps,angle=270,width=8cm}
    \epsfig{file=kebisawa_E1_fig5_4.eps,angle=270,width=8cm}
  \end{center}
\caption{Flare-like light curves of some of the variable sources
with soft spectra.}  
\label{lightcurve}
\end{figure*}

\begin{acknowledgements}
The observation was carried out under the Chandra guest observer program by NASA.

\end{acknowledgements}

\end{document}